\documentclass{article} \usepackage{epsfig,latexsym,graphics}
\usepackage{amsmath} \usepackage{graphicx,epstopdf}
\def\circa#1{\,\raise.3ex\hbox{$#1$\kern-.75em\lower1ex\hbox{$\sim$}}\,}

   \newcommand{\be}{\begin{equation}}
\newcommand{\ee}{\end{equation}} \newcommand{\ben}{\begin{displaymath}}
\newcommand{\een}{\end{displaymath}} \newcommand{\ba}{\begin{eqnarray}}
\newcommand{\ea}{\end{eqnarray}} \newcommand{\ban}{\begin{eqnarray*}}
\newcommand{\ean}{\end{eqnarray*}}

\def\eeeq{\end{eqnarray}} 

\def\beeq{\begin{eqnarray}} \def\eeeq{\end{eqnarray}} 
  \def\to{\rightarrow}

  \def\ID{1 \kern -.45 em 1}

\begin{document}\vspace{1.cm}
{\centering {\Large\bf Infrared weak corrections to strongly interacting
gauge bosons scattering}

\vspace{1.cm} { \large{\bf Paolo
Ciafaloni}\footnote{paolo.ciafaloni@le.infn.it} and {\bf Alfredo
Urbano}\footnote{alfredo.urbano@le.infn.it}}\\ {\it INFN - Sezione di Lecce
and Universit\`a del Salento}\\ \centerline{\it Via per Arnesano, I-73100
Lecce, Italy }
\vspace{0.4cm} }
\vspace{0.3cm}
\begin{abstract}
We evaluate the impact of electroweak corrections of infrared origin on
longitudinal strongly interacting gauge bosons scattering, calculating all
order resummed expressions at the double log level. As a working example,
we consider the Standard model with a heavy Higgs. At energies typical of
forthcoming experiments (LHC,ILC,CLIC), the corrections are in the 10-40\%
range, the relative sign depending on the initial state considered and on
whether or not additional gauge bosons emission is included.
\end{abstract}

\section{Introduction}
Hopefully, experiments at the Large Hadron Collider (LHC) will soon unveil
the mystery surrounding the way in which the SU(2) $\otimes$ U(1)
electroweak (EW) symmetry is implemented in fundamental physics. The
mechanism of elementary particles masses generation, the profound reason
why the EW symmetry is apparently respected by the interactions but broken
by the spectrum of the theory, are all puzzles that have been awaiting for
a clear answer for a long time, and will be most probably be clarified
soon. In the framework of the Standard Model (SM), if the Higgs particle is
light, i.e. close to the present experimental upper bound of 114.4 GeV
\cite{pdg}, the Higgs sector, which also includes the Goldstone bosons, is
characterized by a perturbative coupling of the same order of the gauge
couplings. Indeed, indications coming from present experimental data at the
100 GeV scale and below seem to favor a light SM Higgs situation of this
kind, or a scenario similar to this one, like supersymmetric extensions of
the SM itself. However, a scenario in which the sector responsible for
Symmetry breaking is a strongly interacting one is not excluded, and many
models of this kind have been considered also in recent times
\cite{strongHiggs}. In this case the scattering of longitudinal gauge
bosons, related to the Goldstone bosons interactions via the equivalence
theorem, is strongly enhanced and provides a direct experimental probe of
the physics that is responsible for symmetry breaking.  The purpose of this
paper is to evaluate the impact of EW radiative corrections of infrared
(IR) origin on the scattering of longitudinal gauge bosons.  As a prototype
of a strongly interacting symmetry breaking sector, we consider the
Standard Model with a heavy Higgs; a more general study will be considered
elsewhere \cite{futuro}.

There are several reasons why studying the impact of EW corrections of IR
origin to strongly interacting longitudinal gauge bosons scattering appears
to be a sensible idea. In first place, these corrections depend on the
exchange of quantas of energy $M\ll \omega\ll \sqrt{s}$, $M$ being the IR
cutoff scale ($M\sim M_W\sim M_Z$) and $\sqrt{s}$ the physical process
scale, of the order of 1 TeV. Then, it is reasonable to expect that
corrections of IR origin are somehow universal, depending only on the known
"low energy" SM physics and not on the unknown "high energy" strong
dynamics at the TeV scale. In second place, these corrections have been
shown in the literature to be significant at the TeV scale: at one loop the
presence of double logs of IR origin produces relative corrections that can
be as big as 30-40\%\cite{exclEW}. Then, it might turn out to be important
to include them into the analysis of longitudinal gauge bosons scattering.

In the present work we mainly consider ``fully inclusive'' observables,
i.e. observables that include gauge bosons emissions in the final state,
and we resum EW corrections at the Double Log level. It is by now an
established fact that, despite na\"{\i}ve
expectations, even fully inclusive EW observables are affected by DLs of IR
origin \cite{inclEW}.  We also consider exclusive observables, affected by
Sudakov logs, for a final comparison.  Since we consider the Higgs mass to
be heavy, of the order of 1 TeV, we cannot directly rely on the results
obtained in the "recovered SU(2) $\otimes $ U(1) symmetry limit"
\cite{exclEW,inclEW},
 where all
energies in the physical processes are considered to be much larger than
the particles masses.  We are then led to consider two different
situations: $M\ll\sqrt{s}< M_H$, and $\sqrt{s}\gg M_H>M_W$, that we
describe in some detail below. The case $\sqrt{s}\gg M_H>M_W$ was
considered in \cite{bosfus}, where fermion-antifermion production was
analyzed. Here we analyze longitudinal gauge bosons scattering
and we also consider energies smaller than the Higgs mass.

\section{The case $\sqrt{s}\ll M_H$}

If the c.m. energy of the process is much smaller than the Higgs mass,
$\sqrt{s}\ll M_H$, the SU(2) $\otimes$ U(1) gauge is of no direct use. In
fact, this symmetry is badly broken in the mass spectrum by the heavy Higgs
mass, and since the Higgs $h$ transforms into the Goldstone bosons
$\varphi_k$, say, under an infinitesimal $SU_L(2)$ isospin
transformation\footnote{In the Appendix the symbols appearing here and in
the following are defined, and a discussion of the symmetry proprieties of
the Lagrangian is given.}:
\begin{equation}\label{gaugetrasformation}
  h\rightarrow h-\frac{1}{2}\varphi_{k}\delta_{kl}\alpha_{l}^{L} \\ \\
  \qquad\qquad \varphi_{k}\rightarrow
  \varphi_{k}+\frac{1}{2}h\alpha_{k}^{L}+
  \frac{1}{2}\varphi_{l}\varepsilon_{lmk}\alpha_{m}^{L},
\end{equation}
where $\alpha_{k=1,2,3}^{L}$ are the parameters of the transformation; then
even the tree level hard cross sections are not related by isospin
symmetry. This does not mean that a calculation of IR effects cannot be
performed of course, but a straightforward calculation of resummed effects
is out of question. Here we prefer to consider the $g'\to 0$ limit, in
which the {\sl custodial} symmetry under which the Higgs transforms as a
singlet and the Goldstones as a triplet is valid also after symmetry
breaking:
 \begin{equation}\label{custodialtransformation}
  h\rightarrow h\\ \\ \qquad\qquad \varphi_{k}\rightarrow \varphi_{k}+
\varphi_{k}+i\alpha_{m}\left(T_{V}^{m}\right)_{kl}\varphi_{l},
\end{equation}
where $\left(T_{V}^{m}\right)_{kl}\equiv i\varepsilon_{kml}$ are the
custodial symmetry generators in the adjoint $SU(2)_{V}$ representation.

Let us first investigate what are the relations on cross sections dictated
by custodial symmetry. Since all our quantities are inclusive over final
states, the cross sections only depend on the initial legs through the so
called overlap matrix \cite{inclEW} (see fig. \ref{overlappa}):
\begin{equation}\label{crosssection}
\mathcal{O}_{\beta_{1},\alpha_{1};\beta_{2},\alpha_{2}}=\langle\beta_{1},\beta_{2}|\mathcal{O}|\alpha_{1},\alpha_{2}\rangle\quad
\mathcal{O}=S^{\dag}S \qquad
d\sigma_{\alpha_{1},\alpha_{2}}=\mathcal{O}_{\alpha_{1},\alpha_{1};\alpha_{2},\alpha_{2}}.
\end{equation}

Because of t-channel custodial symmetry invariance, we have:

\begin{equation}\label{WEtheorem}
\left[\overrightarrow{T}_{V},\mathcal{O}^{H}\right]=0\,\,\Rightarrow\,\,\langle
T_{V}^{\prime},T_{V}^{3\,\prime}|\mathcal{O}^{H}|
T_{V},T_{V}^{3}\rangle=C_{T_{V}}^{H}\delta_{T_{V}T_{V}^{\prime}}\delta_{T_{V}^{3}T_{V}^{3\,\prime}},
\end{equation}
where, following the signs convention reported in fig. \ref{overlappa}, we
have defined the total custodial generators on the t-channel that coupling
legs $1$ and $1^{\prime}$ as $T^{a}_{V}\equiv
T^{a}_{V,1}-T^{a}_{V,1^{\prime}}$ and where the action of the isospin
generators on the overlap matrix is described by the relations:
\begin{equation}\label{azionesullaoverlap}
\left(T^{a}_{V,1}\mathcal{O}^{H}\right)_{\beta_{1},\alpha_{1};\beta_{2},\alpha_{2}}=\sum_{\delta_{1}}\left(T^{a}_{V,1}\right)_{\alpha_{1}\delta_{1}}
\mathcal{O}^{H}_{\beta_{1},\delta_{1};\beta_{2},\alpha_{2}}
\end{equation}
and:
\begin{equation}
\left(T^{a}_{V,1'}\mathcal{O}^{H}\right)_{\beta_{1},\alpha_{1};\beta_{2},\alpha_{2}}=\sum_{\delta_{1}}\left(T^{a}_{V,1'}\right)_{\beta_{1}\delta_{1}}
\mathcal{O}^{H}_{\delta_{1},\alpha_{1};\beta_{2},\alpha_{2}}.
\end{equation}
Using the states:
\begin{eqnarray}\label{ciliegia} 
  |2,0\rangle &=&
  \frac{1}{\sqrt{6}}\left(|\varphi^{-}\varphi^{+}\rangle-2|\varphi_{3}\varphi_{3}\rangle+|\varphi^{+}\varphi^{-}\rangle\right),
  \\ |1,0\rangle&=&
  \frac{1}{\sqrt{2}}\left(|\varphi^{+}\varphi^{-}\rangle-|\varphi^{-}\varphi^{+}\rangle\right),
  \\ |0,0\rangle&=&
  \frac{1}{\sqrt{3}}\left(|\varphi^{+}\varphi^{-}\rangle+|\varphi^{-}\varphi^{+}\rangle+|\varphi_{3}\varphi_{3}\rangle\right)\label{muso}
\end{eqnarray}
and specializing eq. (\ref{WEtheorem}) to
\begin{equation}\label{sviluppo}
\langle1,0|\mathcal{O}|0,0\rangle=0\,\,\,\,\,\,\langle2,0|\mathcal{O}|1,0\rangle=0\,\,\,\,\,\,\langle0,0|\mathcal{O}|2,0\rangle=0,
\end{equation}
we obtain a system whose solutions are the $SU(2)_V$ constraints on the
cross sections:
\begin{eqnarray}
\sigma_{++} &=& \sigma_{--}\label{gazzaladra}\\ \sigma_{3+} &=& \sigma_{3-}
\\ \sigma_{33} &=& \sigma_{++}+\sigma_{-+}-\sigma_{3+}\,.\label{cesaroni}
\end{eqnarray}
These constraints are satisfied by the hard cross sections and by the
dressed ones.\\ The hard cross sections, in the $g'\to 0$ limit, are:
\begin{equation}\label{scaduta1}
\sigma_{++}^{H}=\frac{1}{32\pi
s}\left\{\frac{g^{2}}{4}\left[\frac{2s+t}{t}+\frac{t-s}{s+t}
\right]-\frac{s}{v^{2}}\right\}^{2},
\end{equation}
\begin{equation*}
\end{equation*}
\begin{equation}\label{scaduta2}
\sigma_{3+}^{H}=\frac{1}{32\pi
s}\left\{\frac{g^{2}}{4}\left[\frac{t-s}{t+s}-\frac{2t+s}{s}\right]+\frac{t}{v^{2}}\right\}^{2}+\frac{1}{32\pi
s}
\left\{\frac{g^{2}}{4}\left[\frac{2s+t}{t}+\frac{2t+s}{s}\right]-\frac{t+s}{v^{2}}\right\}^{2},
\end{equation}
\begin{equation*}
\end{equation*}
\begin{equation}\label{scaduta3}
\sigma_{33}^{H}=\frac{1}{16\pi
s}\left\{\frac{g^{2}}{4}\left[\frac{s-t}{s+t}-\frac{2s+t}{t}\right]+\frac{s}{v^{2}}\right\}^{2},
\end{equation}
\begin{equation*}
\end{equation*}
\begin{align}\label{scaduta4}
\sigma_{-+}^{H}=& \frac{1}{32\pi
s}\left\{\frac{g^{2}}{4}\left[\frac{2s+t}{t}+\frac{2t+s}{s}\right]-\frac{t+s}{v^{2}}\right\}^{2}+
\frac{1}{32\pi
s}\left\{\frac{g^{2}}{4}\left[\frac{s-t}{s+t}+\frac{2t+s}{s}\right]-\frac{t}{v^{2}}\right\}^{2}\nonumber\\&+
\frac{1}{32\pi
s}\left\{\frac{g^{2}}{4}\left[\frac{s-t}{s+t}-\frac{2s+t}{t}\right]+\frac{s}{v^{2}}\right\}^{2},
\end{align}
where $s,t,u$ are the Mandelstam variables as usually defined.
The expression of the eikonal current that describes the emission of a soft
gauge boson $W^{a}$ from the external legs of the overlap matrix is given
by:
\begin{equation}\label{correntediemissione}
J_V^{a,\mu}(k)=g\left[\frac{p_{1}^{\mu}}{2p_{1}\cdot
k}\left(T_{V,1}^{a}-T_{V,1'}^{a}\right)+\frac{p_{2}^{\mu}}{2p_{2}\cdot
k}\left(T_{V,2}^{a}-T_{V,2'}^{a}\right)\right];
\end{equation}
squaring this current, summing over all the possible gauge bosons emitted
$a$, we obtain the following insertion operator written in the Feynman
gauge:
\begin{equation}\label{insertionoperator}
\mathcal{I}(k)=g^{2}\frac{2p_{1}\cdot p_{2}}{(2p_{1}\cdot k)(2p_{2}\cdot
k)}\left(\overrightarrow{T_{V}}_{,1}-
\overrightarrow{T_{V}}_{,1'}\right)\cdot\left(\overrightarrow{T_{V}}_{,2}-
\overrightarrow{T_{V}}_{,2'}\right).
\end{equation}
Because of the conservation of the total custodial generator on the
t-channel, we have:
\begin{equation}\label{insertionoperator2}
\mathcal{I}(k)=-g^{2}\frac{2p_{1}\cdot p_{2}}{(2p_{1}\cdot k)(2p_{2}\cdot
k)}|\overrightarrow{T_{V}}|^{2}
\end{equation}
and the resummed expression for the overlap matrix is:
\begin{equation}\label{resummedexpression}
\mathcal{O}=e^{S(s,M_{W})}\mathcal{O}^{H}=
\exp\left[-\frac{\textbf{L}_{W}}{4}|\overrightarrow{T_{V}}|^{2}\right]\mathcal{O}^{H},
\end{equation}
where:
\begin{equation}\label{elle}
\textbf{L}_{W}=\frac{g^{2}}{2}\int\frac{d^{3}\overrightarrow{k}}{(2\pi)^{3}2\omega_{k}}\frac{2p_{1}\cdot
p_{2}}{(p_{1}\cdot k)(p_{2}\cdot
k)}=\frac{\alpha_{w}}{4\pi}\ln^{2}\frac{s}{M_{W}^{2}}.\,\,\,\,\,\,\,\,\,\,\,\,\,\,\,(\alpha_{w}=g^{2}/4\pi)
\end{equation}
At this point it's straightforward to convert
eq. (\ref{resummedexpression}) into a system of equations that are able to
connect the electroweak corrected cross sections and the tree level
ones. In fact we just have to use the states classified in
eqns. (\ref{ciliegia}$\div$\ref{muso}), considering the overlap matrix
elements between them; a simple analysis allow us to write:
\begin{equation}\label{allows}
\langle 1,0|\mathcal{O}|1,0\rangle=\langle
1,0|\exp\left[-\frac{T_{V}(T_{V}+1)\textbf{L}_{W}}{4}\right]\mathcal{O}^{H}
|1,0\rangle=\exp\left[-\frac{\textbf{L}_{W}}{2}\right]\langle
1,0|\mathcal{O}^{H}|1,0\rangle;
\end{equation}
Using eq. (\ref{ciliegia}) and the relations between the overlap matrix
elements and the usual cross sections, this equation becomes:
\begin{equation}\label{primaequazione}
\sigma_{+-}-\sigma_{++}=e^{-\textbf{L}_{W}/2}\left(\sigma^{H}_{+-}-\sigma^{H}_{++}\right);
\end{equation}
reasoning in a similar way for the states $|2,0\rangle$ and $|0,0\rangle$
we can obtain two other fundamental relations:
\begin{equation}\label{secondaequazione}
\sigma_{+-}+\sigma_{++}-3\sigma_{3+}+2\sigma_{33}=e^{-3\textbf{L}_{W}/2}
\left(\sigma_{+-}^{H}+\sigma_{++}^{H}-3\sigma_{3+}^{H}+2\sigma_{33}^{H}\right)
\end{equation}
and
\begin{equation}\label{terzaequazione}
2\sigma_{+-}+2\sigma_{++}+4\sigma_{3+}+\sigma_{33}=2\sigma_{+-}^{H}+2\sigma^{H}_{++}+4\sigma^{H}_{3+}+\sigma^{H}_{33}.
\end{equation}
Solving the system obtained from
eqns. (\ref{primaequazione}$\div$\ref{terzaequazione}) we can write the
following expressions for dressed independent cross sections:
\begin{align}
\sigma_{\pm+}=&\sigma_{++}^{H}\left(\frac{1}{3}\pm\frac{1}{2}e^{-\frac{1}{2}\textbf{L}_{W}}+\frac{1}{6}e^{-\frac{3}{2}\textbf{L}_{W}}\right)+
\sigma_{3+}^{H}\left(\frac{1}{3}-\frac{1}{3}e^{-\frac{3}{2}\textbf{L}_{W}}\right)\label{zucc1}\\
&+
\sigma_{-+}^{H}\left(\frac{1}{3}\mp\frac{1}{2}e^{-\frac{1}{2}\textbf{L}_{W}}+\frac{1}{6}e^{-\frac{3}{2}\textbf{L}_{W}}\right),\nonumber
\end{align}
\begin{equation}\label{zucc2}
\sigma_{3+}=\sigma_{++}^{H}\left(\frac{1}{3}-\frac{1}{3}e^{-\frac{3}{2}\textbf{L}_{W}}\right)+
\sigma_{-+}^{H}\left(\frac{1}{3}-\frac{1}{3}e^{-\frac{3}{2}\textbf{L}_{W}}\right)+
\sigma_{3+}^{H}\left(\frac{1}{3}+\frac{2}{3}e^{-\frac{3}{2}\textbf{L}_{W}}\right)
\end{equation}
and, for completeness, the expression for $\sigma_{33}$ that can be
obtained from (\ref{cesaroni}):
\begin{equation}\label{risci}
\sigma_{33}=(\sigma_{-+}^{H}+\sigma_{++}^{H})\left(\frac{1}{3}+\frac{2}{3}e^{-\frac{3}{2}\textbf{L}_{W}}\right)+
\sigma_{3+}^{H}\left(\frac{1}{3}-\frac{4}{3}e^{-\frac{3}{2}\textbf{L}_{W}}\right).
\end{equation}

\section{The case $\sqrt{s}\gg M_H>M_{W}$}

If the c.m. energy of the process is much higher than the Higgs mass,
$\sqrt{s}\gg M_H>M_{W}$, in the limit $g^{\prime}\to 0$ the custodial
symmetry is extended to a global $SU(2)_{L}\otimes SU(2)_{R}$ symmetry (see
Appendix).  In order to identify the most useful way in which we can use
this symmetry considering the resummation of the electroweak corrections,
it's necessary to take a look at the form of the eikonal current in this
energy region; when $\sqrt{s}\gg M_H>M_{W}$ the situation is complicated by
the fact that, after the gauge boson emission, one can have as final state
a gauge boson as well as an Higgs particle. Considering the usual eikonal
approximation and the equivalence theorem it's possible to construct the
current for the emission of a soft gauge boson of momentum $k$, Lorentz
index $\mu$ and isospin index $b$ out of a longitudinal gauge boson of
isospin index $a$ and momentum $p$:
\begin{equation}\label{masscutoffs}
       J^{\mu}_{ac}(k,b;p)=\frac{gp^{\mu}}{2p\cdot k}\left[i
   \varepsilon_{abc}\Theta_{W}-i\delta_{ab}\Theta_{H}\right],
\end{equation}
where $\Theta_{W}\equiv\vartheta\left(2p\cdot k-M_{W}^{2}\right)$ and
$\Theta_{H}\equiv\vartheta\left(2p\cdot k-M_{H}^{2}\right)$ are the usual
Heaviside functions and where the second term on the right hand side take
into account the presence of the Higgs into the final state; we can choose
to use a more compact and useful matrix notation, as follows\footnote{In
the following discussion we use the notation $T^{a=1,2,3}_{V,L,R}$
referring to the contribution of a single leg, without any other additional
index $i$.}:
\begin{equation}\label{morecompactnotation}
   J^{\mu}(k,b;p)=\frac{gp^{\mu}}{2p\cdot k}\left[
   T^{b}_{V}\Theta_{W}+T^{b}_{H}\Theta_{H}\right]= \frac{gp^{\mu}}{p\cdot
   k}\left[
   T^{b}_{L}\Theta_{H}+\frac{1}{2}T^{b}_{V}\left(\Theta_{W}-\Theta_{H}\right)\right],
\end{equation}
in the basis $\left(\varphi_{1},\varphi_{2},\varphi_{3},h\right)^{T}$
where:
\begin{equation}\label{generoigeneratori}
\left(T^{b}_{V}\right)_{ac}=\left(
                            \begin{array}{cc}
                              i\varepsilon_{abc} & 0 \\ 0 & 0 \\
                            \end{array}
                          \right) \qquad \left(T^{b}_{H}\right)_{ac}=\left(
                            \begin{array}{cc}
                            0 & -i\delta_{ab} \\ i\delta_{bc} & 0 \\
                            \end{array}
                          \right) \qquad
T_{L}=\frac{1}{2}\left(T_{V}+T_{H}\right) \qquad
T_{R}=\frac{1}{2}\left(T_{V}-T_{H}\right).
\end{equation}
In eq. (\ref{masscutoffs}) when the energy of the emitted boson $\omega$ is
such that $M_{H}<\omega<M_{W}$, the Higgs boson contribution is turned off
and the current is the same as the one achieved in the previous paragraph;
when $\omega>M_{H}$, contrarily, we have to consider also an Higgs
additional contribution that, as we can see in (\ref{morecompactnotation}),
affects the $T^{a}_{V}$ term and forces to introduce a further one
proportional to $T^{a}_{L}$.\\ The obtained expression for the eikonal
emission current as explicit function of the operators $T^{a=1,2,3}_{L}$
and $T^{a=1,2,3}_{V}$ leads automatically to the correct way in which we
must look to the states in the t-channel; in fact we have to consider the
diagonal subgroup of $SU(2)_{L}\otimes SU(2)_{R}$ generated through
$T^{a}_{V,i}\equiv T^{a}_{L,i}+T^{a}_{R,i}$, on a single leg denoted by
$i$, and then to classify the states according to the quantum numbers of
the total t-channel Casimir $|\overrightarrow{T_{V}}|^{2}$ operator;
considering the notation $|T_{L},T_{R};T_{V}\rangle$, we are left with $6$
physical overlap states:
 \begin{eqnarray} \label{vettorone}
|0,0;0\rangle &=&
\frac{1}{2}\left(-|\varphi^{+}\varphi^{-}\rangle-|hh\rangle-|\varphi_{3}\varphi_{3}\rangle-|\varphi^{-}\varphi^{+}\rangle\right),
\\ |0,1;1\rangle &=&
\frac{1}{2}\left(-|\varphi^{+}\varphi^{-}\rangle+i|\varphi_{3}h\rangle-i|h\varphi_{3}\rangle+|\varphi^{-}\varphi^{+}\rangle\right),
\\ |1,0;1\rangle &=&
\frac{1}{2}\left(-|\varphi^{+}\varphi^{-}\rangle-i|\varphi_{3}h\rangle+i|h\varphi_{3}\rangle+|\varphi^{-}\varphi^{+}\rangle\right),
\\ |1,1;2\rangle &=&
\frac{1}{\sqrt{6}}\left(-2|\varphi_{3}\varphi_{3}\rangle+|\varphi^{+}\varphi^{-}\rangle+|\varphi^{-}\varphi^{+}\rangle\right),
\\ |1,1;1\rangle &=&
\frac{1}{\sqrt{2}}\left(i|h\varphi_{3}\rangle+i|\varphi_{3}h\rangle\right),\\
|1,1;0\rangle &=&
\frac{1}{2\sqrt{3}}\left(3|hh\rangle-|\varphi_{3}\varphi_{3}\rangle-|\varphi^{+}\varphi^{-}\rangle-|\varphi^{-}\varphi^{+}\rangle\right),
 \end{eqnarray}
  matching all to the eigenvalues $T_{V}^{3}=0$. Notice that the
 $SU(2)_{V}$ constraints in eqns. (\ref{gazzaladra}$\div$\ref{cesaroni})
 between the cross sections are still valid; in addiction we shall have
 other relations characteristics of the $SU(2)_{L}\otimes SU(2)_{R}$
 symmetry. Reasoning as in the previous paragraph, through the explicit
 evaluation of the overlap matrix element $\langle
 1,1;0|\mathcal{O}|0,0;0\rangle=0$, we are able to write:
\begin{equation}\label{ulteriorerelazione}
    \sigma_{hh}=\sigma_{++}+\sigma_{+-}+\sigma_{3+}-2\sigma_{h+}.
\end{equation}
The tree level cross sections are:
\begin{equation*}
\end{equation*}
\begin{equation}\label{primabosonica}
\sigma_{++}^{H}=\frac{\pi\alpha_{W}^{2}}{32
s}\left[\frac{2s+t}{t}-\frac{s-t}{t+s}-\frac{2M_{H}^{2}}{M_{W}^{2}}\right]^{2},
\end{equation}
\begin{equation*}
\end{equation*}
\begin{equation}\label{secondabosonica}
\sigma_{33}^{H}=\frac{\pi\alpha_{W}^{2}}{32
s}\left[3\left(\frac{s-t}{s+t}-\frac{2s+t}{t}-\frac{M_{H}^{2}}{M_{W}^{2}}\right)^{2}+9\frac{M_{H}^{4}}{M_{W}^{4}}\right],
\end{equation}
\begin{equation*}
\end{equation*}
\begin{align}
\sigma_{-+}^{H}=& \frac{\pi\alpha_{W}^{2}}{32 s}
\left\{\left(\frac{2t+s}{s}+\frac{2s+t}{t}+\frac{2M_{H}^{2}}{M_{W}^{2}}\right)^{2}+\left(
\frac{s-t}{s+t}-\frac{2t+s}{s}-\frac{2M_{H}^{2}}{M_{W}^{2}}\right)^{2}\right. \\
&\left.+2\left(\frac{s-t}{s+t}-\frac{2s+t}{t}-\frac{M_{H}^{2}}{M_{W}^{2}}\right)^{2}+
2\left(\frac{2t+s}{s}-\frac{s-t}{t+s}-\frac{2s+t}{t}\right)^{2}
\right\},\nonumber
\end{align}
\begin{equation*}
\end{equation*}
\begin{align}
\sigma_{3+}^{H}=&\frac{\pi\alpha_{W}^{2}}{32 s}
\left\{\left(\frac{2t+s}{s}+\frac{s-t}{s+t}+\frac{M_{H}^{2}}{M_{W}^{2}}\right)^{2}+
\left(\frac{2s+t}{t}+\frac{2t+s}{s}-\frac{M_{H}^{2}}{M_{W}^{2}}\right)^{2}
\right.\nonumber\\
&+\left.2\left(\frac{2t+s}{s}-\frac{2s+t}{t}-\frac{s-t}{s+t}\right)^{2}\right\},
\end{align}
in which as usual $\alpha_{W}=\frac{g^{2}}{4\pi}$. Once  the
expression of the eikonal current is known, it's straightforward in the overlap
formalism to obtain the expressions of the dressed cross sections. The
procedure follows closely the one described in the previous section, and we
obtain (see also \cite{bosfus}):
\begin{equation}\label{risommazione}
\mathcal{O}=e^{\mathcal{S}(s,M_{H}^{2},M_{W}^{2})}\mathcal{O}^{H},
\end{equation}
where $\mathcal{S}(s,M_{H}^{2},M_{W}^{2})$ is obtained through an energy
ordered integration of the eikonal factor given by the square of the
emission current described in eq. (\ref{morecompactnotation}):
\begin{equation}\label{primasimmetria}
 S(s,M_{H},M_{W})=\left(-\overrightarrow{T}_{L}^{2}\textbf{L}_{H}-\frac{\overrightarrow{T}_{V}^{2}}{4}
 (\textbf{L}_{W}-\textbf{L}_{H})\right),
\end{equation}
where:
\begin{equation}\label{goldstone}
\textbf{L}_{W}=\frac{\alpha_{W}}{4\pi}\ln^{2}\left(\frac{s}{M_{W}^{2}}\right)\vartheta(\sqrt{s}-M_{W}),
\end{equation}
and
\begin{align}\label{dandy}
\textbf{L}_{H}=&\frac{\alpha_{W}}{\pi}\left[\ln^{2}\left(\frac{\sqrt{s}}{M_{W}}\right)-2\ln^{2}\left(\frac{M_{H}}{M_{W}}\right)\right]
\vartheta\left(\sqrt{s}-\frac{M_{H}}{M_{W}}\right)+\\
&\left[\frac{2\alpha_{W}}{\pi}\ln^{2}\left(\frac{\sqrt{s}}{M_{H}}\right)\right]\vartheta(\sqrt{s}-M_{H})
\vartheta\left(\frac{M_{H}}{M_{W}}-\sqrt{s}\right).\nonumber
\end{align}
Starting from (\ref{risommazione}) the pathway to obtain the explicit
expressions for the dressed cross sections follows how pointed in the
previous paragraph, considering obviously the states classified in
(\ref{vettorone}) as external physical states for the overlap matrix. A
straightforward calculation leads to:
\begin{align}
\sigma_{\pm+}=&\frac{1}{4}\left(\sigma_{++}^{H}+\sigma_{-+}^{H}+2\sigma_{3+}^{H}\right)+\frac{1}{12}
\left(\sigma_{++}^{H}+\sigma_{-+}^{H}-2\sigma_{3+}^{H}\right)e^{-2\textbf{L}_{H}}
\label{eq1}\\&\pm
\frac{1}{4}\left(\sigma_{++}^{H}-\sigma_{-+}^{H}+
2\mathcal{I}^{H}\right)e^{-\frac{1}{2}\left(\textbf{L}_{W}-\textbf{L}_{H}\right)}\pm
\frac{1}{4}\left(\sigma_{++}^{H}-\sigma_{-+}^{H}-
2\mathcal{I}^{H}\right)e^{-2\textbf{L}_{W}+\frac{3}{2}\left(\textbf{L}_{W}-\textbf{L}_{H}\right)}\nonumber\\&+
\frac{1}{6}\left(\sigma_{++}^{H}+\sigma_{-+}^{H}-
2\sigma_{3+}^{H}\right)e^{-2\textbf{L}_{W}+\frac{1}{2}\left(\textbf{L}_{W}-\textbf{L}_{H}\right)}\,,\nonumber
\end{align}
\begin{align}
\label{eq2}\sigma_{3+}=&\frac{1}{4}\left(\sigma_{++}^{H}+\sigma_{-+}^{H}+2\sigma_{3+}^{H}\right)+\frac{1}{12}
\left(\sigma_{++}^{H}+\sigma_{-+}^{H}-2\sigma_{3+}^{H}\right)e^{-2\textbf{L}_{H}}\\&-
\frac{1}{3}\left(\sigma_{++}^{H}+\sigma_{-+}^{H}-2\sigma_{3+}^{H}\right)
e^{-2\textbf{L}_{W}+\frac{1}{2}\left(\textbf{L}_{W}-\textbf{L}_{H}\right)}\,,\nonumber
\end{align}
\begin{align}\label{eq3}
\sigma_{33}=&\frac{1}{4}\left(\sigma_{++}^{H}+\sigma_{-+}^{H}+2\sigma_{3+}^{H}\right)+\frac{1}{12}
\left(\sigma_{++}^{H}+\sigma_{-+}^{H}-2\sigma_{3+}^{H}\right)e^{-2\textbf{L}_{H}}\\&+
\frac{2}{3}\left(\sigma_{++}^{H}+\sigma_{-+}^{H}-2\sigma_{3+}^{H}\right)
e^{-2\textbf{L}_{W}+\frac{1}{2}\left(\textbf{L}_{W}-\textbf{L}_{H}\right)}\,,\nonumber
\end{align}
\begin{align}\label{eq4}
\sigma_{h+}=&\frac{1}{4}\left(\sigma_{++}^{H}+\sigma_{-+}^{H}+2\sigma_{3+}^{H}\right)-\frac{1}{4}\left(\sigma_{++}^{H}+
\sigma_{-+}^{H}-2\sigma_{3+}^{H}\right)e^{-2\textbf{L}_{H}}\,,
\end{align}
where
$\mathcal{I}^{H}=\frac{1}{2}\left(\sigma_{0+}-\sigma_{0^{*}+}\right)$. We
report below also the explicit expressions for these particular
contributions, that are:
\begin{eqnarray}
  \sigma_{0+}^{H} &=& \frac{\pi\alpha_{W}^{2}}{32
  s}\left\{\left(\frac{2s+t}{t}+2\frac{s-t}{s+t}+\frac{M_{H}^{2}}{M_{W}^{2}}\right)^{2}+
  \left(2\frac{2s+t}{t}+\frac{s-t}{s+t}-\frac{M_{H}^{2}}{M_{W}^{2}}\right)^{2}\right\},
  \\ \sigma_{0^{*}+}^{H} &=&\frac{\pi\alpha_{W}^{2}}{32
  s}\left\{\left(\frac{2s+t}{t}-2\frac{2t+s}{s}-\frac{M_{H}^{2}}{M_{W}^{2}}\right)^{2}+
  \left(2\frac{2t+s}{s}-\frac{s-t}{s+t}-\frac{M_{H}^{2}}{M_{W}^{2}}\right)^{2}\right\}.
\end{eqnarray}

From eqs. (\ref{eq1},\ref{eq2},\ref{eq3}) it is easy to obtain the
asymptotic ($\sqrt{s}\to\infty$) behavior of the cross sections. Namely,
some particular combinations between the cross sections are radiative
invariants, that is combinations that are free from the DL corrections;
these invariants are\footnote{notice that in the case at hand a direct
calculation gives $\sigma_{h+}^{H}=\sigma_{3+}^{H}$.}:
\begin{eqnarray}
\Theta_{0} &=&
\frac{\sigma_{++}+\sigma_{-+}+\sigma_{h+}+\sigma_{3+}}{4}\label{gazze}, \\
\Theta_{1} &=& \frac{\sigma_{++}-\sigma_{-+}+2\mathcal{I}}{4},
\end{eqnarray}
  In the limit $\sqrt{s}\rightarrow
\infty$ it's possible to establish a precise asymptotic behaviour of the
electroweak corrected cross sections; in fact in this situation we have:
\begin{equation}\label{limitecitato}
\lim_{\sqrt{s}\rightarrow
\infty}e^{-\left(\textbf{L}_{W}-\textbf{L}_{H}\right)}=e^{-\frac{2\alpha_{W}}{\pi}\ln^{2}\frac{M_{H}}{M_{W}}}
\end{equation}
and:
\begin{equation}\label{altrilimiti}
\lim_{\sqrt{s}\rightarrow
\infty}e^{-\textbf{L}_{W}}=0\,\,\,\,\,\,\,\lim_{\sqrt{s}\rightarrow
\infty}e^{-\textbf{L}_{H}}=0,
\end{equation}
so the cross sections in this limit become:
\begin{eqnarray}
  \sigma_{++} &\rightarrow&
  \frac{1}{4}\left(\sigma_{++}^{H}+\sigma_{-+}^{H}+2\sigma_{3+}^{H}\right)+\frac{1}{4}
  \left(\sigma_{++}^{H}-\sigma_{-+}^{H}\right)e^{-\frac{\alpha_{W}}{\pi}\ln^{2}
\frac{M_{H}}{M_{W}}}\label{mouse},
  \\ \sigma_{-+} &\rightarrow&
  \frac{1}{4}\left(\sigma_{++}^{H}+\sigma_{-+}^{H}+2\sigma_{3+}^{H}\right)-\frac{1}{4}
  \left(\sigma_{++}^{H}-\sigma_{-+}^{H}\right)e^{-\frac{\alpha_{W}}{\pi}\ln^{2}
\frac{M_{H}}{M_{W}}}\label{incidente},
  \\ \sigma_{33,3+} &\rightarrow&
  \frac{1}{4}\left(\sigma_{++}^{H}+\sigma_{-+}^{H}+2\sigma_{3+}^{H}\right)\label{detto}.
\end{eqnarray}
Finally, let us now consider resummed DL EW corrections of infrared origin
for exclusive observables, i.e. observables in which additional gauge
bosons emission is forbidden. In this case, the treatment of Sudakov DLs is
analogous to the known results present in the literature \cite{exclEW},
but we have to
take into account the mass splitting between the weak scale and the Higgs mass.
The resummed cross section is obtained by
multiplying each external leg $i$ by an exponential factor as follows:
\begin{equation}\label{sudakovcorrection}
\sigma^{Sud}=\sigma^{H}\exp\left[-2\sum_{i}\left(t_{L}^{i}(t_{L}^{i}+1)
\textbf{L}_{H}+\frac{t_{V}^{i}(t_{V}^{i}+1)}{4}
(\textbf{L}_{W}-\textbf{L}_{H})\right)\right],
\end{equation}
where, compared with (\ref{primasimmetria}), this last expression shows a
sum over the charged external legs labeled by $i$;
 $t_{L(V)}^i$ is referred to a
single leg and not to a double leg composition.

\section{Graphics and Comments}

\begin{figure}[!H!]
\begin{center}
\includegraphics[width=15cm]{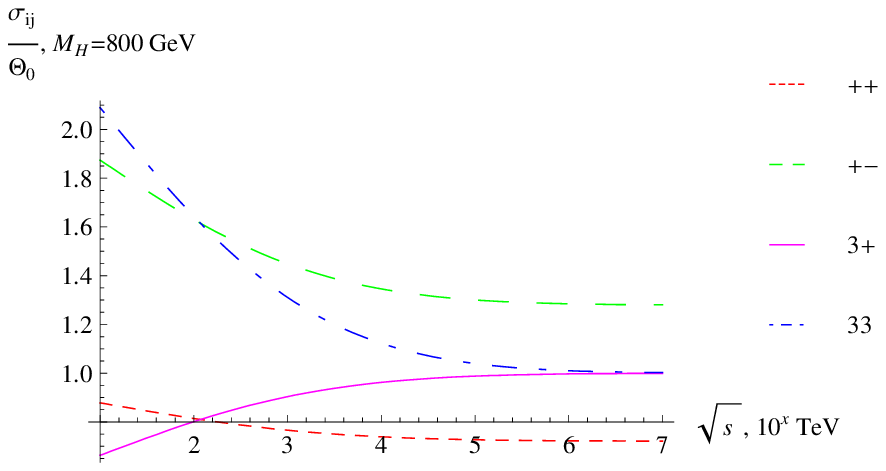}\\
  \caption{Case $\sqrt{s}\gg M_{H}>M_{W}$: ratio between the dressed cross
  sections and the invariant $\Theta_{0}$, as defined in (\ref{gazze}), as
  function of the c.m. energy.}\label{true5bis}
  \end{center}
\end{figure}

\begin{figure}[!H!]
      \begin{center}\label{angtot}
      \includegraphics[width=15cm] {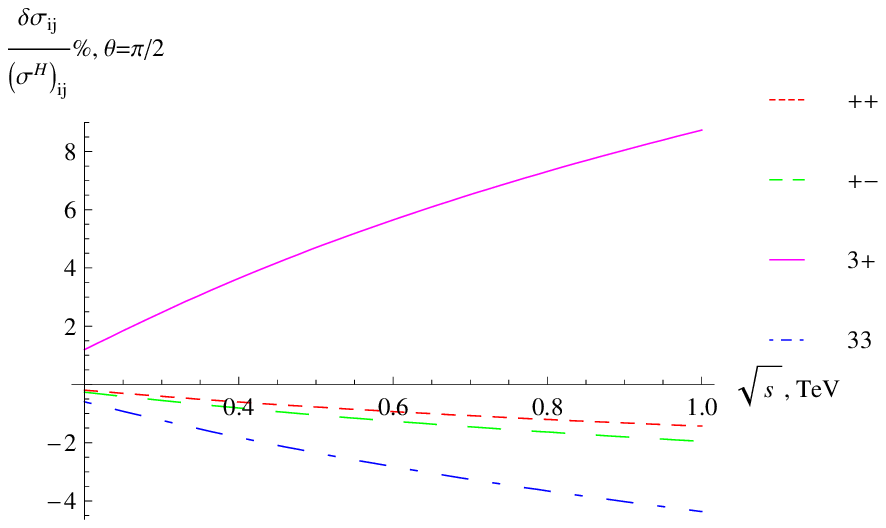}
          \caption{Case $\sqrt{s}\ll M_{H}$: ratio between the dressed
          cross sections and the tree level ones as function of the
          c.m. energy, at fixed scattering angle $\theta=\pi/2$ and for
          $M_{H}=800$ GeV.  }
  \end{center}
\end{figure}

Our results are given in eqs. (\ref{zucc1},\ref{zucc2},\ref{risci}) for the
case $\sqrt{s}<M_H$ and eqs. (\ref{eq1},\ref{eq2},\ref{eq3}) for the case
$\sqrt{s}\gg M_H\gg M_W$: they represent the all-order resummed expression
for EW radiative corrections at the double log level. Here we have
considered fully inclusive observables (i.e., gauge bosons radiation in the
final state is always included); the corresponding tree level cross
sections for longitudinal gauge bosons scattering are given in
eqs. (\ref{scaduta1},\ref{scaduta2},\ref{scaduta3}) for $\sqrt{s}<M_H$ and
eqs.  (\ref{eq1},\ref{eq2},\ref{eq3}) for $\sqrt{s}\gg M_H\gg M_W$. The
asymptotic behavior of the cross sections can be seen in
fig. \ref{true5bis}: for very high energies every single cross section
tends to a value which is a linear combination of the "radiative
invariants" defined in eq. (\ref{gazze}). In this regime radiative
corrections are of the same order of tree level values; notice however that
this situation is valid for energies that are far too high for current or
near future experiments. At the TeV scale relevant for LHC and ILC (1 TeV)
or CLIC (3 TeV) the situation is depicted in fig.  \ref{angtot}. Inclusive
radiative corrections are below 10 \% under the TeV scale, i.e. at the
level of, or bigger than, NLO QCD corrections \cite{QCD}. Relative
corrections
grow towards
the 30\% value as the invariant mass of the scattering gauge bosons reaches
3 TeV.

\begin{figure}[!H!]
\begin{center}
\includegraphics[width=15cm]{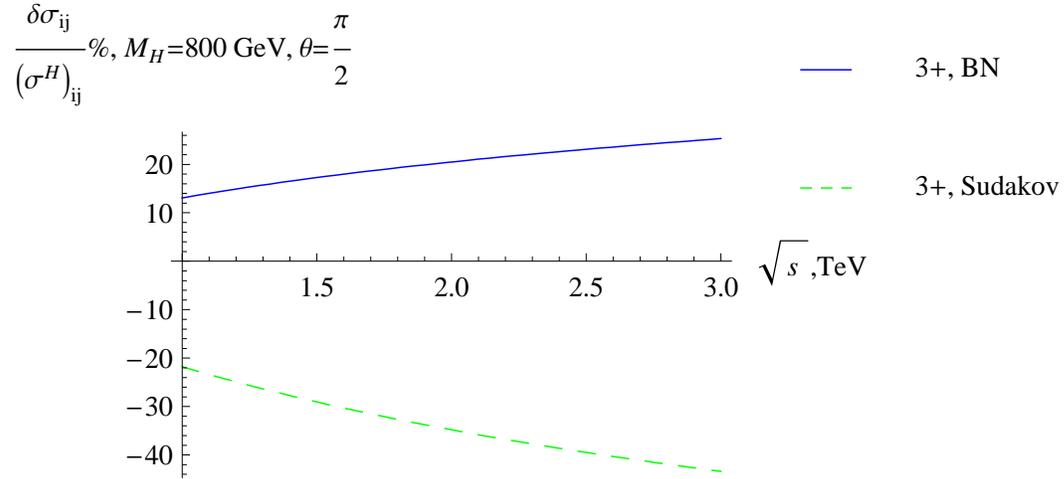}\\
  \caption{Case $\sqrt{s}\gg M_{H}>M_{W}$: Block-Nordsieck corrections
  vs. Sudakov corrections in the $\sigma_{3+}$ case. In the inclusive
  treatment the electroweak corrections change their sign compared to the
  exclusive case.}\label{true6bis}
  \end{center}
\end{figure}

\begin{figure}[!H!]
\begin{center}
\includegraphics[width=15cm]{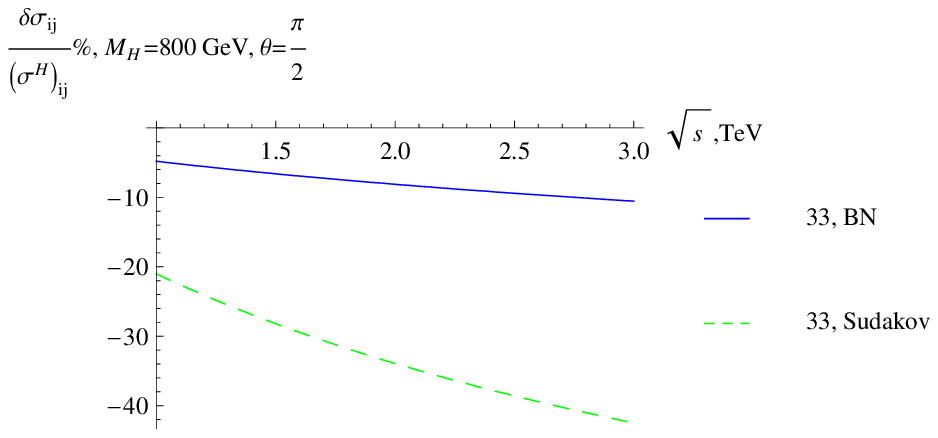}\\
  \caption{Case $\sqrt{s}\gg M_{H}>M_{W}$: Block-Nordsieck corrections
  vs. Sudakov corrections in the $\sigma_{33}$ case. In the inclusive
  treatment the electroweak corrections have the same sign if compared with
  the exclusive case, but the cross section results less
  suppressed.}\label{true7bis}
  \end{center}
\end{figure}

It is particularly interesting to compare the EW corrections of IR origin
for inclusive observables with the ones for exclusive observables
(fig. \ref{true6bis} and \ref{true7bis}). The latter are the ones usually
considered in the literature: it is usually assumed that additional gauge
boson emission is either irrelevant and/or produces a final state that is
distinguishable from the one produced by the hard scattering. However, it
has been noticed for instance in \cite{Baur} that at the LHC, even with the
actual experimental cuts a certain degree of weak boson emission may escape
detection and needs to be included. We plot in fig. \ref{true6bis} and in
fig. \ref{true7bis} the corrections to the fully inclusive cross sections
$\sigma_{3+}$ and $\sigma_{33}$, labeled "BN", and to the exclusive cross
sections, which includes only virtual EW corrections, labeled "Sudakov"
\cite{exclEW}. The case of $\sigma_{3+}$ is particularly interesting since
radiative corrections range between -40 \% and +25 \% depending on the
definition of the observable. The relevant value for EW corrections of
infrared origin will depend on the experimental setup and on the various
cuts defining the observables; we think it is important to notice that in
any case, for the ``standard'' exclusive definition there is a relevant
suppression of the signal. Our result is compatible with the ones obtained
in \cite{LogEW} in view of the different treatments (complete one loop
vs. DL resummed, light vs. heavy Higgs and so on). Finally, in the case of
 $\sigma_{33}$, which is interesting for a linear collider where the two
final electrons are doubly tagged, radiative corrections turn out to be
negative both in the inclusive and exclusive case; however the exclusive
case is more suppressed and reaches the -40\% value at 3 TeV.

Overall, electroweak radiative corrections to strongly interacting
longitudinal bosons turn out to be potentially relevant for next generation
of colliders. Here we have chosen to use the SM heavy Higgs case as a
prototype and we find that these corrections are in the  -40 \% + 25 \%
range for energies below 3 TeV. The impact of these corrections on the
analysis of the EW symmetry breaking sector through boson scattering
depends on the model considered and on the details of the experimental
cuts; we postpone a more refined study to a subsequent paper \cite{futuro}.

\begin{figure}[!H!]
\begin{center}
 \includegraphics[width=4cm]{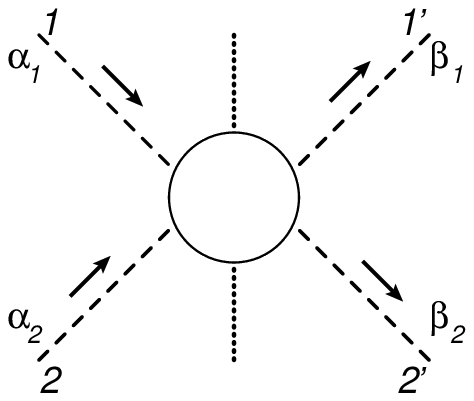} \qquad\qquad\qquad
\includegraphics[width=10cm]{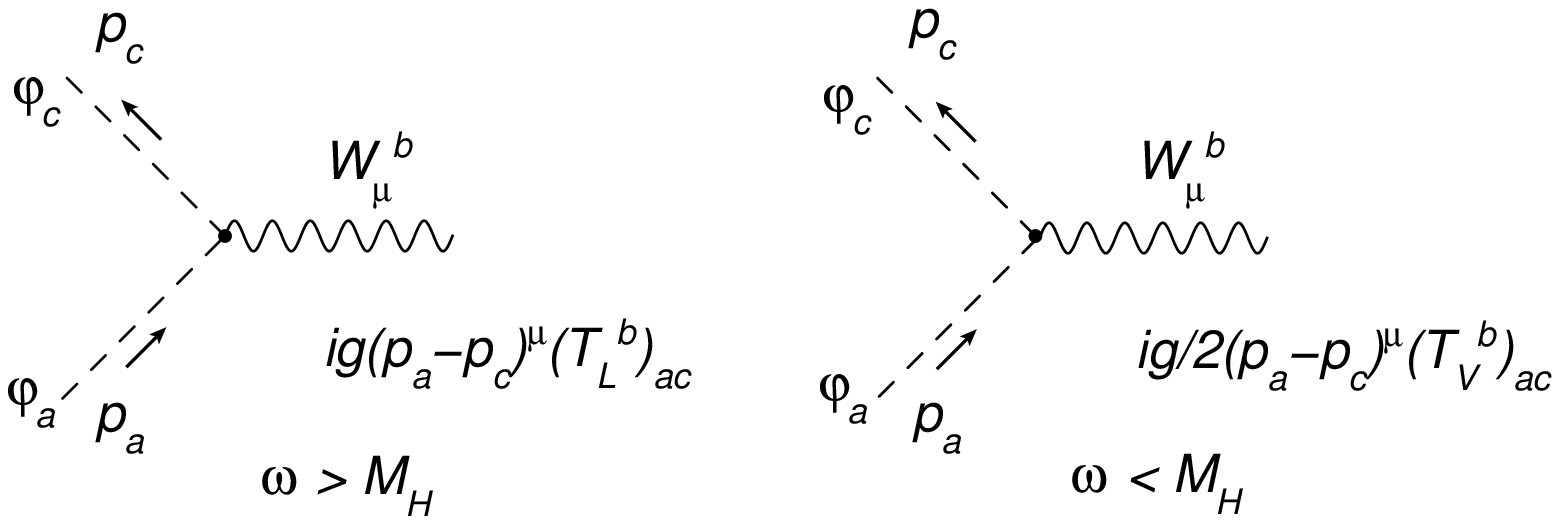}
\caption{Left: Diagrammatic representation of the overlap matrix. The
  t-channel couples legs $1$ and $1^{\prime}$, while the choice of the
  momentum sign on the external legs fixes the conservation law for the
  generators of the considered symmetry as:
  $T_{1}^{a}-T_{1'}^{a}=T_{2'}^{a}-T_{2}^{a}$.  Right: Feynman rules for
  the scalar-scalar-gauge interactions from (\ref{lagrangianadigauge}) in
  the two situations $\omega >M_{H}$ and $\omega <M_{H}$.  In the first
  case we have chosen the basis in which $\varphi_{4}\equiv h$.  $T_{L}$
  and $T_{V}$ are those defined in (\ref{generigeneratori}).
  }\label{overlappa}
  \end{center}
\end{figure}

\begin{figure}[!H!]
      \begin{center}
      \includegraphics[width=15cm] {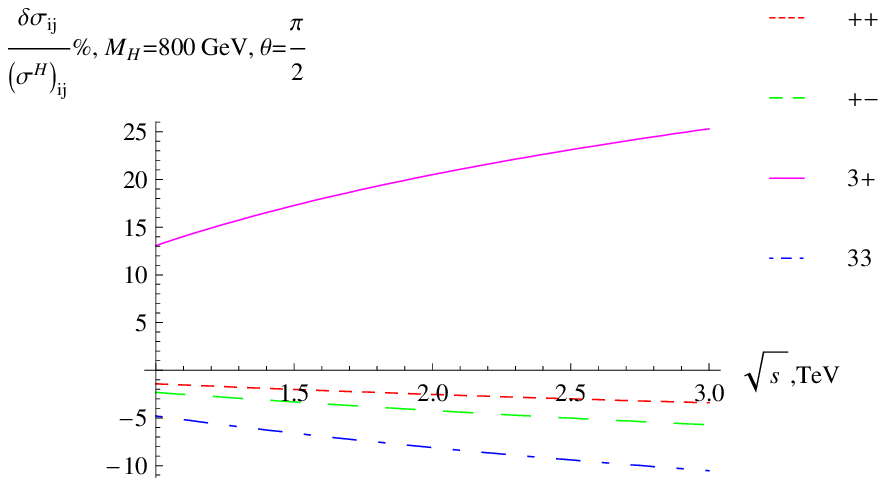}
          \caption{Case $\sqrt{s}\gg M_{H}>M_{W}$: ratio between the
          dressed cross sections and the tree level ones as function of the
          c.m. energy, at fixed scattering angle $\theta=\pi/2$ and for
          $M_{H}=800$ GeV.  }\label{angtot2}
  \end{center}
\end{figure}


\appendix

\section{Appendix}

In this appendix we illustrate the main symmetry proprieties of the
 Standard Model Lagrangian in the limit $g^{\prime}\to 0$, taking into
 account in particular the role of the gauge symmetry and of its global
 $SU(2)_{L}\otimes SU(2)_{R}$ extension.\\ Since we work considering just
 the scalar sector, our starting point is the Lagrangian of the $SU(2)$
 gauged $\sigma$-model:
\begin{equation}\label{bandabassotti}
    \mathcal{L}=-\frac{1}{2}Tr\left(\widehat{W}_{\mu\nu}\widehat{W}^{\mu\nu}\right)+\frac{1}{2}Tr\left[(D_{\mu}\Phi)^{\dag}D^{\mu}\Phi\right]
    -\frac{\mu^{2}}{2}Tr\left(\Phi^{\dag}\Phi\right)-\frac{\lambda}{4}\left[
    Tr\left(\Phi^{\dag}\Phi\right)\right] ^{2},
\end{equation}
where the scalar content of the theory is organized into the following
matrix:
\begin{equation}\label{riorganizzo}
\Phi\equiv
\frac{1}{\sqrt{2}}\left(h\textbf{1}_{2\text{x}2}+i\varphi_{i}\tau_{i}\right)=
\frac{1}{\sqrt{2}} \left(
  \begin{array}{cc}
    h+i\varphi_{3} & \varphi_{1}-i\varphi_{2} \\ \varphi_{1}+i\varphi_{2} &
    h-i\varphi_{3} \\
  \end{array}
\right)
\end{equation}
and where, as usual:
\begin{equation}\label{definitio}
\begin{array}{c}
  \widehat{W}_{\mu}\equiv \frac{1}{2}W_{\mu}^{k}\tau_{k}, \\ \\
   \widehat{W}_{\mu\nu}=\partial_{\mu}\widehat{W}_{\nu}-\partial_{\nu}\widehat{W}_{\mu}+ig\left[\widehat{W}_{\mu},\widehat{W}_{\nu}\right],\\
   \\ D_{\mu}\Phi=\partial_{\mu}\Phi+ig\widehat{W}_{\mu}\Phi.
\end{array}
\end{equation}
Taking into account this notation, in which the matrix $\Phi$ acquires the
standard ``\emph{doublet-antidoublet}" form, the gauge transformations,
restricted to the isospin $SU(2)_{L}$ case in the limit $g^{\prime}\to 0$,
are:
\begin{equation}\label{gaugetrasfo}
    \begin{array}{c}
      ig\widehat{W}_{\mu}^{\prime}=g_{L}(x)ig\widehat{W}_{\mu}g_{L}^{\dag}(x)+g_{L}(x)\partial_{\mu}g_{L}^{\dag}(x),
      \\ \\
      \widehat{W}_{\mu\nu}^{\prime}=g_{L}(x)\widehat{W}_{\mu\nu}g_{L}^{\dag},\\
      \\ \Phi^{\prime}=g_{L}(x)\Phi,
    \end{array}
\end{equation}
where:
\begin{equation}\label{singolarmente}
g_{L}(x)=\exp\left(i\alpha_{k}(x)\frac{\tau_{k}}{2}\right)\in SU(2)_{L}.
\end{equation}
In order to clarify the transformation proprieties under the gauge
symmetry, it's straightforward to consider explicitly (\ref{gaugetrasfo})
for the scalar fields, obtaining:
\begin{equation}\label{trafocagauge}\left\{
\begin{array}{c}
  h\rightarrow h-\frac{1}{2}\varphi_{k}\delta_{kl}\alpha_{l}^{L}(x), \\ \\
   \varphi_{k}\rightarrow \varphi_{k}+\frac{1}{2}h\alpha_{k}^{L}(x)+
   \frac{1}{2}\varepsilon_{klm}\varphi_{l}\alpha_{m}^{L}(x).
\end{array}\right.
\end{equation}
At this point it's possible to see that the Lagrangian in
(\ref{bandabassotti}), because of the limit $g^{\prime}\to 0$, has a larger
global $SU(2)_{L}\otimes SU(2)_{R}$ symmetry, under which the scalar fields
transform as:
\begin{equation}\label{enhanced}
\Phi^{\prime}=L\Phi R^{\dag},
\end{equation}
with $L\in SU(2)_{L}$ and $R\in SU(2)_{R}$ or, expanding the fields as in
(\ref{riorganizzo}):
\begin{equation}\label{trafoca}\left\{
\begin{array}{c}
  h\rightarrow
   h+\frac{1}{2}\varphi_{k}\delta_{kl}\left(\alpha_{l}^{R}-\alpha_{l}^{L}\right)
   \\ \\ \varphi_{k}\rightarrow
   \varphi_{k}+\frac{1}{2}h\left(\alpha_{k}^{L}-\alpha_{k}^{R}\right)+
   \frac{1}{2}\varepsilon_{lmk}\varphi_{l}\left(\alpha_{m}^{R}+\alpha_{m}^{L}\right)
\end{array}\right.
\end{equation}
After electroweak symmetry breaking the global $SU(2)_{L}\otimes SU(2)_{R}$
symmetry is spontaneously broken into its diagonal custodial subgroup
$SU(2)_{V}$; the transformation laws under this custodial symmetry for the
scalar fields are:
 \begin{equation}\label{trafoca2}\left\{
\begin{array}{c}
  h\rightarrow h\\ \\ \varphi_{k}\rightarrow
   \varphi_{k}+\varepsilon_{lmk}\varphi_{l}\alpha_{m}=
   \varphi_{k}+i\left(T_{V}^{m}\right)_{kl}\alpha_{m}\varphi_{l}.
\end{array}\right.
\end{equation}
As a consequence the Higgs boson $h$ is a singlet, while the three
Goldstone bosons are a triplet, transforming according to the adjoint
representation of $SU(2)_{V}$.\\ Once assumed these transformation
proprieties it's possible to consider the explicit interactions in the
Lagrangian (\ref{bandabassotti}) that we have used during our work.\\
Introducing the Higgs mass as function of the parameters in the scalar
potential as $M_{H}^{2}=2\lambda v^{2}$, the interactions that involve the
scalar fields are described by:
\begin{equation}\label{lagrangianainterazioniscalari}
\mathcal{L}=-\frac{M_{H}^{2}}{8v^{2}}h^{4}-\frac{M_{H}^{2}}{2v}h^{3}-\frac{M_{H}^{2}}{4v^{2}}h^{2}\varphi_{k}\varphi_{l}\delta_{kl}
-\frac{M_{H}^{2}}{2v}h\varphi_{k}\varphi_{l}\delta_{kl}-\frac{M_{H}^{2}}{8v^{2}}\left(\varphi_{k}\varphi_{l}\delta_{kl}\right)
\left(\varphi_{m}\varphi_{n}\delta_{mn}\right),
\end{equation}
while the gauge interactions of the scalar fields, written in the
$g^{\prime}\to 0$ limit, are:
\begin{equation}\label{lagrangianadigauge}
\mathcal{L}=\frac{g}{2}\delta_{kl}W_{\mu}^{l}\left[h\left(\partial^{\mu}\varphi_{k}\right)-\varphi_{k}\left(\partial^{\mu}h\right)\right]+
\frac{g}{2}\varepsilon_{klm}\varphi_{k}W_{\mu}^{l}\left(\partial^{\mu}\varphi_{m}\right)+\frac{g^{2}v}{4}g^{\mu\nu}hW_{\mu}^{k}W_{\nu}^{l}\delta_{kl}.
\end{equation}
Considering in particular the $3$-vertex interactions $\mathcal{L}(h\varphi
W, \varphi\varphi W)$ of the previous Lagrangian, as in
fig.\ref{overlappa}, it's possible to construct the eikonal current that
describes the emission of a soft gauge boson from a longitudinal one, with
the possibility to have, as final state, a gauge boson as well as an Higgs
boson,
 obtaining the current:
\begin{equation}
   J^{\mu}(k,b;p)=\frac{gp^{\mu}}{2p\cdot k}\left[
   T^{b}_{V}\Theta_{W}+T^{b}_{H}\Theta_{H}\right]= \frac{gp^{\mu}}{p\cdot
   k}\left[
   T^{b}_{L}\Theta_{H}+\frac{1}{2}T^{b}_{V}\left(\Theta_{W}-\Theta_{H}\right)\right],
\end{equation}
with:
\begin{equation}\label{generigeneratori}
\left(T^{b}_{L}\right)_{ac}=\frac{1}{2}\left(
                            \begin{array}{cc}
                              i\varepsilon_{abc} & -i\delta_{ab} \\
                              i\delta_{bc} & 0 \\
                            \end{array}
                          \right)=\frac{1}{2}\left(T^{a}_{V}+T^{a}_{H}\right)\,\,\,\,\,\,\,\,\,\,\,\,\,\,\,\,
                          \left(T^{b}_{R}\right)_{ac}\equiv\frac{1}{2}\left(
                            \begin{array}{cc}
                              i\varepsilon_{abc} & +i\delta_{ab} \\
                              -i\delta_{bc} & 0 \\
                            \end{array}
                          \right)=\frac{1}{2}\left(T^{a}_{V}-T^{a}_{H}\right);
\end{equation}
at this point the proprieties of symmetry follow directly from the
commutation relations:
\begin{equation}\label{commutazione}
\left[T^{a}_{V},T^{b}_{V}\right]=i\varepsilon_{abc}T^{c}_{V}\,\,\,\,\,\,\,\,\,\,\,
\left[T^{a}_{H},T^{b}_{H}\right]=i\varepsilon_{abc}T^{c}_{H}\,\,\,\,\,\,\,\,\,\,\,
\left[T^{a}_{H},T^{b}_{H}\right]=i\varepsilon_{abc}T^{c}_{V},
\end{equation}
and:
\begin{equation}\label{commutazione2}
\left[T^{a}_{L},T^{b}_{L}\right]=i\varepsilon_{abc}T^{c}_{L}\,\,\,\,\,\,\,\,\,\,\,
\left[T^{a}_{R},T^{b}_{R}\right]=i\varepsilon_{abc}T^{c}_{R}\,\,\,\,\,\,\,\,\,\,\,
\left[T^{a}_{L},T^{b}_{R}\right]=0,
\end{equation}
from which $\left\{T^{a=1,2,3}_{V},T^{a=1,2,3}_{H}\right\}$ are the six
generators of the $O(4)$ group and
$\left\{T^{a=1,2,3}_{L},T^{a=1,2,3}_{R}\right\}$ are the generators of the
$SU(2)_{L}\otimes SU(2)_{R}\sim O(4)$ group; as a consequence,
$\left\{T^{a=1,2,3}_{V}\right\}$ are the generators of the custodial
$SU(2)_{V}$ diagonal subgroup.

\end{document}